\documentclass[aps,prl,reprint]{revtex4-1}
\usepackage{natbib}
\usepackage{rotating,graphicx} \usepackage{amsmath} \usepackage{amsfonts}
\usepackage{amssymb} \usepackage{color} \usepackage{textcomp} \usepackage{bm}
\usepackage{float} 
\usepackage{hyperref}

\ProvideTextCommand{\DJ}{OT1}{\raisebox{0.25ex}{-}\kern-0.4em D}
\newcommand{\Dbar}{\raisebox{0.25ex}{-}\kern-0.4em D}

\begin{document} 

\title{Chain Length Dispersity Effects on Mobility of Entangled Polymers}
\author{Brandon L. Peters,$^1$ K. Michael Salerno,$^2$ Ting Ge,$^3$
Dvora Perahia,$^4$ and Gary S. Grest$^1$}
\affiliation{$^1$Sandia National Laboratories, Albuquerque, New Mexico
87185, USA}
\affiliation{$^2$U. S. Naval Research Laboratory, Washington, DC 20375,
USA}
\affiliation{$^3$Department of Mechanical Engineering and Materials Science, Duke University, Durham, North Carolina 27708, USA}
\affiliation{$^4$Department of Chemistry and Department of Physics and
Astronomy, Clemson University, Clemson, South Carolina 29634, USA}

\date{\today}

\begin{abstract}
While nearly all theoretical and computational studies of entangled polymer melts have focused on uniform samples, polymer synthesis routes always result in some dispersity, albeit narrow, of distribution of molecular weights ($\Dbar_M=M_w/M_n$ $\sim$ 1.02-1.04).  
Here the effects of dispersity on chain mobility are studied for entangled, disperse melts using a coarse-grained model for polyethylene. Polymer melts with chain lengths set to follow a Schulz-Zimm distribution for the same average $M_w = 36$ kg/mol with $\Dbar_M$ = $1.0$ to $1.16$, were studied for times of $600-800$ $\mu$s using molecular dynamics simulations. This time frame is longer than the time required to reach the diffusive regime. We find that dispersity in this range does not affect the entanglement time or tube diameter. However, while there is negligible difference in the average mobility of chains for the uniform distribution $\Dbar_M=1.0$ and $\Dbar_M = 1.02$, the shortest chains move significantly faster than the longest ones offering a constraint release pathway for the melts for larger $\Dbar$. \end{abstract}
\maketitle

The dynamics of macromolecules drive the unique viscoelastic properties that underline their strength and flexibility. Polymer chains consist of a large number of atoms that often exceeds $10^6$, constituting entropic objects whose properties scale with their molecular weight. Variations in molecular weights, or the dispersity, have an immense effect on their phase behavior and dynamics and consequently affect numerous technologies, particularly those that incorporate entangled polymers with controlled elasticity.

The variability in molecular weights stems from inherent polymerization synthesis routes that yield dispersity in polymer chain length. This dispersity is a result of the statistical process that determines the polymerization path which is manifested in the differences of the number average molecular weight $M_n$ and the weight average molecular weight $M_w$ \cite{rubinstein}.  The degree of dispersity $\Dbar_M$ is defined as the ratio of $M_w/M_n$ \cite{gilbert.2009}. Among the lowest dispersity polymers are those made by anionic and atom-transfer radical polymerization \cite{matyjaszewski2001atom,chiefari1998living}, which exhibit relatively narrow distributions $\Dbar_M \sim 1.02-1.04$. The dispersity of these polymers is well captured by the Schulz-Zimm distribution \cite{rubinstein,zimm1948scattering,hiemenz.2007}.  
Seemingly a small number, this variability in chain lengths reflects a wide distribution of molecular weights where the ratio of the shortest to the longest chain length for the Schulz-Zimm distribution is three even for $\Dbar_M =1.02$. This corresponds to a difference in relaxation times of 27 assuming a standard reptation exponent of 3.0, larger if one uses the experimentally observed value of 3.4. The effect of systematically varying $\Dbar_M$ on the dynamics of entangled melts is not easily accessible experimentally and remains an open question theoretically notwithstanding immense effort \cite{graessley1967viscosity,graessley1986effects,schieber1986kinetic,doi1987dynamics,rubinstein1988self,tsenoglou1991molecular,milner1996relating,wang2003relaxation,leonardi2000rheological,wagner2001molecular,cassagnau1993rheology,descloizeaux1988}. 
Most of these studies has focused on blending mixtures of two chain lengths. Here with the power of new computational tools, we
address the effects of narrow distributions of $M_w$ within the framework anionic and atom-transfer radical polymerization on 
chain mobility in entangled melts. This fundamental aspect of polymer physics has not been thoroughly explored, and the understanding of the constraint release pathways in which dispersity affects polymer response remains an open question.  Consistent with earlier dynamic theories and simulations \cite{rubinstein,dorgan2013molecular, descloizeaux1988}, our molecular dynamics (MD) simulations show that the presence of highly mobile short chains leads to constraint release for longer chains.

The significance of polymeric mechanical response has resulted in thorough efforts to resolve the effects of the dispersity of molecular weights on the flow of entangled melts. 
The flow characteristics are often captured in terms of the dependence of the viscosity on shear rate and linear viscoelastic
response which are sensitive to chain dispersity. The effects become particularity significant for high molecular weights \cite{struglinski1985,wasserman1992}.
Dispersed melts have been treated theoretically by extending models of melts with uniform chain length \cite{rubinstein,doi1988theory}. These models focused on linear viscoelasticity of entangled polymer melts \cite{graessley1986effects,schieber1986kinetic,doi1987dynamics,rubinstein1988self,tsenoglou1991molecular,milner1996relating,wang2003relaxation,leonardi2000rheological,cassagnau1993rheology}.

These theories clearly demonstrate that the dynamics of linear chains in dispersed polymeric melts cannot be described by the classical reptation theory.  
Only models which explicitly consider the effects of the disperse surroundings of a chain through tube renewal can describe the dispersity effects on observed rheological response. While essentially all previous theoretical work on dispersed polymer melts have focused on linear viscoelastic response, few have discussed the effect of polymer dispersity on chain mobility that underlines the viscoelastic response 
\cite{rubinstein1988self}.  Molecular dynamic simulations allow us to study dispersed entangled polymer melts, bridging the gap between average behavior captured by viscoelastic theories and chain mobility.

Numerical simulations are optimally positioned to study chain mobility in disperse melts. Previous numerical studies
of disperse polymers melts have largely focused on binary blends \cite{baschnagel1998statics,masubuchi2008comparison,martinez2005lattice,barsky2000molecular,picu2007coarse,read2012full,barsky1999nonequilibrium}. However, due to computational limitations, only the longer of the two chain lengths was well above the entanglement molecular weight $M_e$. There have been few studies of polymer melts with a distribution of chains lengths, though mostly for short, unentangled polymers \cite{harmandaris.1998,mavrantzas1999end,baschnagel1998statics,pant1995variable,daoulas2003variable,rorrer2014effects,rorrer2014finding,dorgan2013molecular,rorrer2014molecular,li.2016}.
Rorrer \emph{et al.} \cite{dorgan2013molecular,rorrer2014molecular,rorrer2014effects,rorrer2014finding} mapped a distribution of chain lengths on a small number of chain lengths and showed that for the same weight-averaged molecular weight, increasing the dispersity in chain lengths gives a lower Rouse time and introduces a broadening of the transition to reptation of the chains. Li \emph{et al.} \cite{li.2016} have shown that even very large dispersity has little effect on the polymer glass transition.

With the significance of understanding the dispersity of polymers on chain mobility in entangles melts, this study has used polyethylene (PE), a well-studied macromolecule, as a model system. Computationally, coarse-grained (CG) models with 3-48 methylene groups per CG pseudo-atom \cite{Padding.2001,Padding.2002,Fukunaga.2002,Ashbaugh.2005,Chen.2006,Salerno.2016,Salerno.2016b} have been developed, providing an essential tool to probe a sufficiently large melt that will allow the distinction of dispersity effects. 
Using a CG model for PE with four methylene groups per CG bead \cite{Salerno.2016,Salerno.2016b,peters2017coarse}, we examine chain mobility of disperse entangled polymer melts with dispersity $\Dbar_M$ in the range of the best synthetic routes and compare the results to a uniform polymer melt. This CG model was chosen since for more than 5 methylene groups per CG bead, one has to include extra beads or other constraints to avoid chains cutting through each other \cite{Padding.2001,Padding.2002,Salerno.2016,Salerno.2016b}. The CG PE model used here has previously derived from fully atomistic simulations \cite{Salerno.2016,Salerno.2016b}. The nonbonded and bonded potentials were determined using an iterative Boltzmann inversion method. Additional details of the methodology can be found in Salerno et al. \cite{Salerno.2016,Salerno.2016b}.
Melts with an average molecular weight $M_w \sim 36$ kg/mol ($\sim 640$ CG beads) for dispersity $\Dbar_M$=1.0, 1.02, 1.04, 1.08, and 1.16 were studied. Using this CG models, we could reach times of order $800$ $\mu$s.  In comparison with earlier simulations of broader dispersity, we focus exclusively on low dispersity to understand its effects on chain mobility.

\begin{figure}[ht]
\includegraphics[width=\linewidth]{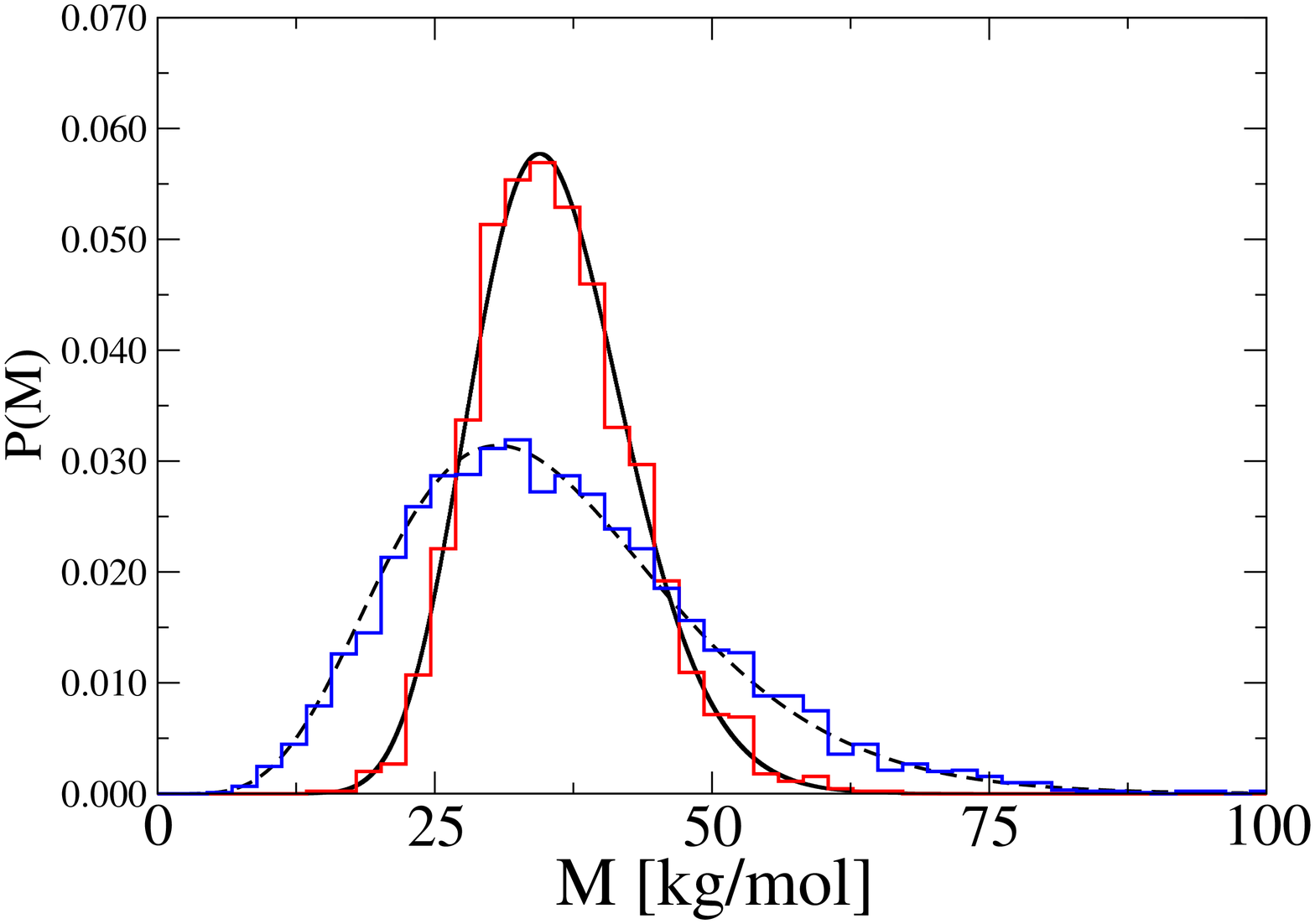}
\caption[orientations]{Distribution $P(M)$ versus molecular weight $M$ for $\Dbar_M$=1.04 (red) for $2000$ chains and $1.16$ (blue) for $4000$ chains compared to the analytic Schulz-Zimm formula (eq.~1).}
\label{fig:dispersity_104_116} 

\end{figure}
To model narrow molecular weight distributions in polymer melts synthesized by anionic and atom-transfer radical polymerization, the chain lengths were chosen to follow a Schulz-Zimm distribution \cite{rubinstein,zimm1948scattering,hiemenz.2007},
\begin{equation} P(M) = \frac{z^{z+1}}{\Gamma (z+1)}\frac{M^{z-1}}{M_n^z} exp \left(\frac{-zM}{M_n} \right)\end{equation}

\noindent where $\Dbar_M=M_w/M_n=(z+1)/z$ \cite{hiemenz.2007}. This distribution captures well the experimental observed molecular weight dispersity as resolved by chromatography \cite{rubinstein}. All systems had the same weight-average molecular weight $M_w = 35.8 \pm 0.2$ kg/mol. Experimentally the entanglement molecular weight $M_e\sim 1.1$-$1.2$ kg/mol \cite{fetters.1999,vega.2004} or about 20 CG beads for PE. Here we use $N_c = 2000$ chains for $1.02\le\Dbar_M \le 1.08$ and 4000 chains for $\Dbar_M =1.16$ to represent the Schulz-Zimm distribution, as shown in Figure 1. For $\Dbar_M=1.0$, $N_c=800$. For the largest dispersity $\Dbar_M =1.16$, the shortest chain ($M/M_e \sim 10$) is well entangled. Within these distributions, there are 398, 524, 671, and 940 unique chain lengths for $\Dbar_M$= 1.02, 1.04, 1.08, and 1.16 respectively. Details of the systems studied here are listed in Table 1. 

\begin{table*}
\caption{Dispersity $\Dbar_M$, number of chains $N_c$, number of CG beads $N_t$, average molecular weight $M_w$, number average molecular weight $M_n$, calculated $\Dbar^c_m=M_w/M_n$, the length of the run $T_r$ (scaled time), number averaged diffusion constant $D$, and weight averaged diffusion constant $\bar{D}$. }
\smallskip
\begin{tabular}{|c|c|c|c|c|c|c|c|c|}
  \hline
  $\Dbar_M$  &$N_c$ & $N_t$ &  $M_w$ [kg/mol] & $M_n$ [kg/mol] & $\Dbar_{m}^c$   & $T_r$ [$\mu$s]& D x $10^{13}$ [m$^2$/s] & $\bar{D}$ x $10^{13}$ [m$^2$/s]\\
  \hline
   1.0  &  800 &  512000   & 35.8 & 35.8 & 1.0   & 790 & 1.15 & 1.15 \\
   1.02  & 2000 & 1251799  & 35.8 & 35.2 & 1.019 & 830 & 1.16 & 1.22 \\
   1.04  & 2000 & 1237358  & 35.6 & 34.4 & 1.036 & 800 & 1.23 & 1.29\\
   1.08  & 2000 & 1176236  & 35.8 & 32.9 & 1.087 & 680 & 1.39 & 1.53 \\
   1.16  & 4000 & 2203172  & 36.0 & 30.8 & 1.169 & 600 & 1.77 & 2.04\\

    \hline
\end{tabular}
\end{table*}

The simulations were performed using the Large Atomic Molecular Massive Parallel Simulator (LAMMPS) molecular dynamics code \cite{Plimpton.1995}. The melts were constructed following the procedure outlined in Auhl \emph{et al.} \cite{auhl2003equilibration} with periodic boundary conditions in all three directions. The simulation was performed at constant volume with the velocity-Verlet integrator and a Langevin thermostat with a damping constant of $100$ ps to maintain the temperature at $500$ K and a time step of $20$ fs. Coarse graining reduces the number of degrees of freedom in a system, creating a smoother free-energy landscape compared with fully atomistic simulations. This results in faster dynamics for the CG polymer chain than for the fully atomistic model \cite{Maranas.2011,Lyubimov.2013,Harmandaris.2009a,Fritz.2011}. For the model used here with 4 methylene groups/CG bead, the dynamic scaling factor $\alpha = 6.2$ \cite{Salerno.2016,Salerno.2016b,peters2017coarse} at $500$ K and at an experimentally relevant density, $\rho = 0.76 g/cm^3$\cite{Fetters.1999.2}. For all the results presented here, time is scaled by $\alpha$, and all five systems were run for $5.0$-$6.6\times 10^9$ time steps. These run times are equivalent to $600$-$800$ $\mu$s and are listed in Table 1 for each system. 

\begin{figure}[ht]
\includegraphics[width=0.4\textwidth]{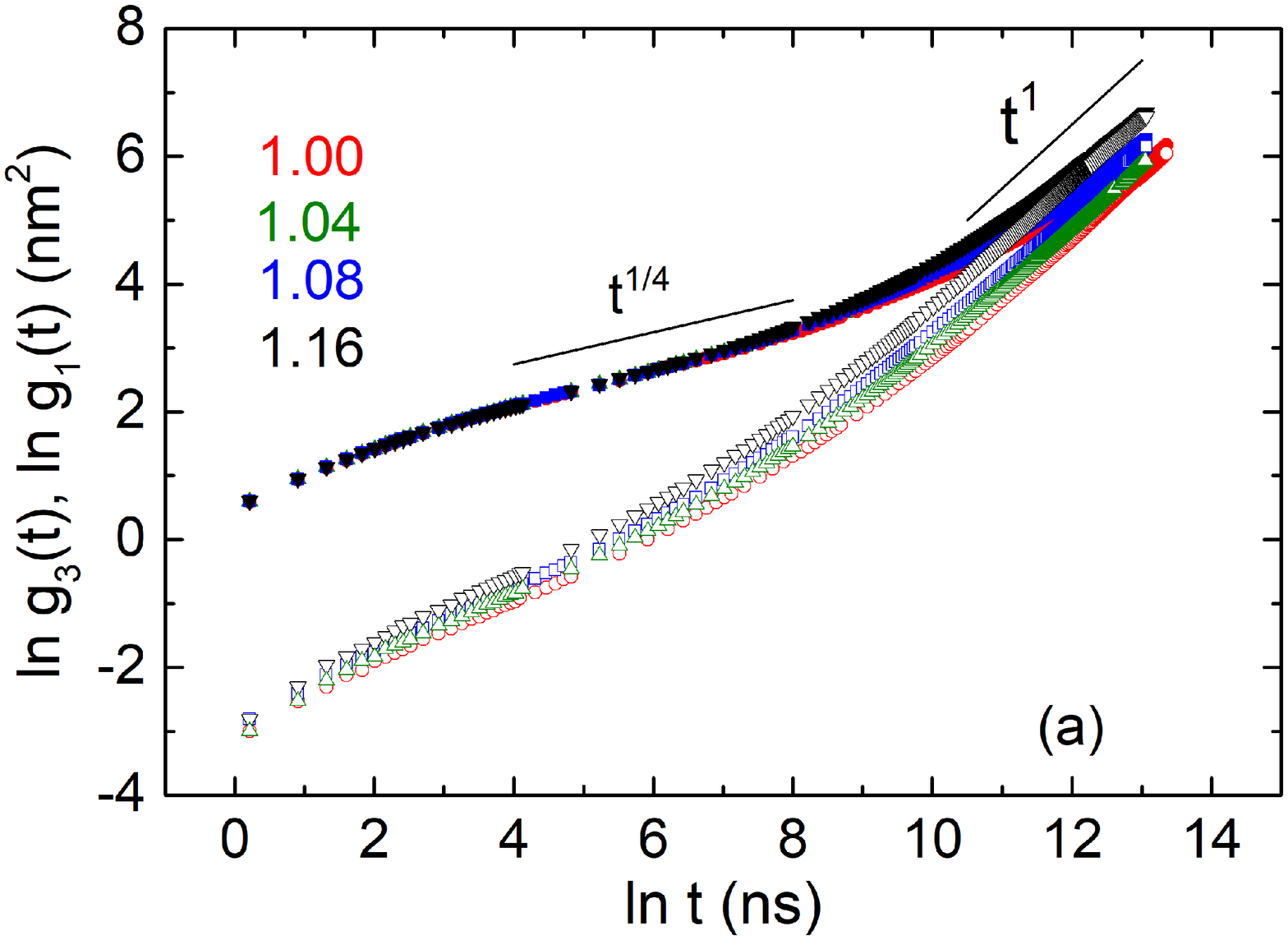}
\includegraphics[width=0.4\textwidth]{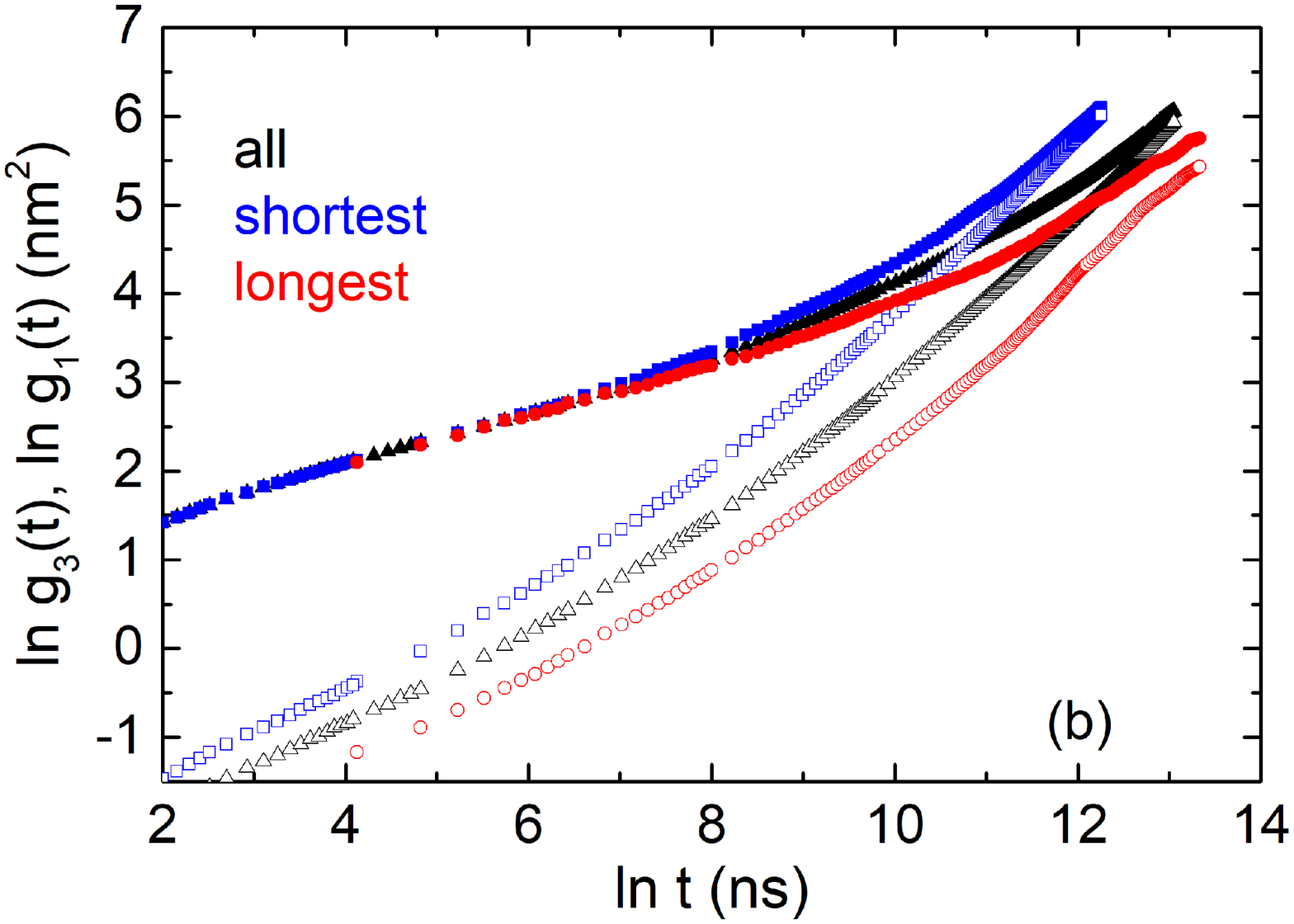}
\caption[orientations]{Mean squared displacement of the center
of mass $g_3(t)$ (open) and center four CG beads $g_1(t)$ (solid) 
for (a) 4 values of the molar mass dispersity ($\Dbar_M$) and (b)
for $\Dbar_M$=1.04, number averaged over all chains (green triangles), the shortest $5\%$ of the chains (blue squares) and longest $5\%$ of the chains (red circles)} \label{fig:MSD}
\end{figure}

The mean-squared displacements (MSD) of the center of mass (cm)
$g_3(t) = \langle (r_{cm}(t) - r_{cm}(0))^2 \rangle$ and the center four CG beads of the chain $g_1(t) = \langle(r_i(t) - r_i(0))^2\rangle$ are shown in Figure 2a for four values of $\Dbar_M$.  The data shown in Figure 2a are averaged over all chains in the system. The two quantities allow the distinction of local motions at short times and macroscopic motion at long times. For long times, the average chain mobility increases and the terminal time $\tau_d$, when the MSDs become diffusive, decreases as $\Dbar_M$ increases from $1.0$ to $1.16$. As seen from the results for the weight averaged diffusion constant $D = g_3(t)/6t$ for $t > \tau_d$ listed in Table 1, $D$ increases by 50\% over the range of $\Dbar_M$ studied.
Results for a uniform melt ($\Dbar_M=1.0$) and the lowest dispersity $\Dbar_M =1.02$ are nearly indistinguishable.  As seen from $g_1(t)$, the motion of the inner monomers at early times does not depend on $\Dbar_M$, as all the chains, even for $\Dbar_M =1.16$, are much longer than the entanglement length. From $g_1(t)$, we find that the crossover from the early Rouse relaxation $t^{1/2}$ regime to the $t^{1/4}$ reptation regime at which topological constraints set in, is at $t^*_e \sim 14$ ns. Assuming that the distribution of segment displacement along the tube is Gaussian on the scale of the tube diameter $d_T$ \cite{hou2017note}, one can determine the entanglement time $\tau_e$ from $t^*_e$ = $\frac{\pi}{9}\tau_e$. This gives $\tau_e \sim 40$ ns.  The MSD of the center monomers at the crossover \cite{hou2017note} $g^*_{1e} = \frac{2}{3\pi} d^2_T$ gives a tube diameter $d_T \sim 4.8$ nm.  Fits to the tube model of the dynamic structure function $S(q,t)$ from neutron spin-echo experiments by Richter \emph{et al. }\cite{Richter.1992} and Schleger \emph{et al.} \cite{Schleger1998} for PE of the same $M_w$ at $509$ K give $\tau_e \sim 5$ ns and $d_T\sim 4.35$ nm.

\begin{figure}[hbt]
\includegraphics[width=\linewidth]{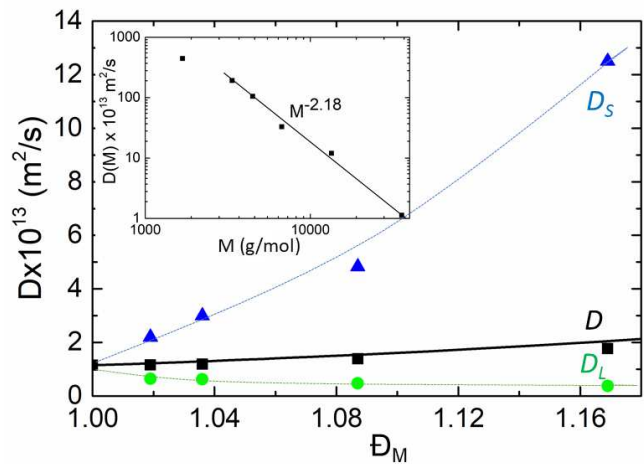}
\caption[orientations]{Diffusion constant $D$ as a function of $\Dbar_M$.  Weight average $D$ averaged over all the chains (black squares), the shortest $5\%$ of the chains $D_S$(blue triangles), and the longest $5\%$ of the chains $D_L$ (green circles).  Blue and green lines are guides to the eye. Error bars are size of symbols. Also shown is $\bar{D}$ calculated from the diffusion constant for uniform melts from eq. 2 (black line). Inset shows the diffusion constant $D(M)$ for uniform melts of molecular weight $M$ and the power law fit for large $M$ given in the text.} \label{fig:D_PDI_scaled}
\end{figure}

The effect of the dispersity is captured through measurements of the mobility of the shortest and longest chains in each melt.
Figure 2b presents results for the MSD of the shortest and longest $5\%$ of the chains compared to the average MSD for all chains for $\Dbar_M=1.04$.  While there is little difference in the motion of the center beads of the chain at early times, at later times the mobility of the beads for the shortest $5\%$ of the chains deviate more from the mean than do the longest $5\%$ of the chains. For $\Dbar_M=1.04$, for the shortest $5\%$ of the chains $M_w=22.0$ kg/mol while $M_w=50.3$ kg/mol for the longest $5\%$, compared to the number average molecular weight of the system $M_w=35.6$ kg/mol. The diffusion constants directly extracted from the simulations are presented in Figure 3, averaged over all chains $D$, the shortest $5\%$ $D_S$ and longest $5\%$ $D_L$.  These results show that the shorter chains move significantly faster than the average and as the dispersity increases, $D_S$ deviates considerably more from $D$ than does $D_L$. 
For comparison, for $\Dbar_M=1.04$, $D(22.0)=3.4$ x $10^{-13}$ m$^2$/s for a uniform melt of chains with $M_w=22.0$ kg/mol is $6\%$ larger than $D_S=3.2$ x $10^{-13}$ m$^2$/s, whereas $D(50.3) = 5.5$ x $10^{-14}$ m$^2$/s for a uniform melt of chains with $50.3$ kg/mol is 13$\%$ smaller than $D_L= 6.3$ x $10^{-14}$ m$^2$/s. These results for $D(M)$ are obtained from simulations for uniform systems presented in the inset of Figure 3 \cite{Grest.2018}.  The ratio of $D$ for a uniform melt of the same $M_w$ as the shortest chains and $D_S$ increases as $\Dbar$ increases whereas the ratio between $D$ for a uniform melt of long chains and $D_L$ does not. For $\Dbar_M=1.16$, $D(11.0)/D_s = 1.2$, while $D(62.8)/D_L = 0.9$, where $M_w=11.0$ kg/mol is molecular weight of the shortest $5\%$ of the chains and $M_w=62.8$ kg/mol is the molecular weight for the longest $5\%$ of the chains for $\Dbar_M=1.16$. 
We also measured the static structure factor $S(q)$ for the entire melt and for the shortest and longest $5\%$ of the chains.  These measurements show no evidence of phase separation of the chains for all $\Dbar$ studied. 
The divergence of the motion of the shortest and longest chains suggest that the short chain enable a constraint release mechanism for the dynamics \cite{doi1988theory,rubinstein}, which is the disentanglement of a chain due to other polymers reptating away.

The diffusion constant of the disperse melts is estimated from the diffusion constant $D(M)$ of uniform melts by incorporating the distribution $P(M)$ of the chains using

\begin{equation}
\bar{D}=\frac{\int D(M) P(M) M dM} { \int P(M) M dM}
\end{equation}

From a series of simulations of uniform polymer melts for $1.6$ kg/mol $\le M \le 35.8$ kg/mol, shown in inset of Figure 3, we find that for large $M$, $D(M)$ is well described by a power law $D(M) = D_1 (M/M_1)^{-2.18}$, where $D_1= 2.81$ x $10^{-10}$ m$^2$/s and $M_1 = 1$ kg/mol \cite{Grest.2018}.  The decay of $D(M)$ with a power law exponent greater than 2 for large $M$ is consistent with experimental results \cite{Lodge.1999}.  Using this power law for $D(M)$ and the Schulz-Zimm distribution (eq.\ 1) for $P(M)$, we estimate $\bar{D}$ using eq.\ 2. As seen from the inset in Figure 3 and Table 1, $\bar{D} > D$ for all $\Dbar_M$. For small $M$, $\bar{D}$ gives a very good estimate of the measured diffusion constant $D$ as the center of the distribution $P(M)$ dominates. However as $M$ increases, the two begin to diverge as the local environment that a chain in uniform melt begins to deviate from that in the dispersed melt. 

This study probed directly the mobility of dispersed entangled polymer melts with distribution as narrow as experimentally attainable for long entangled polymers.  Overall the average mobility of the chains increases as the dispersity increases. We observe that though the average mobility is hardly affected within this dispersity range, the mobility of the shortest and longest chains deviates considerably from the average. The increase diffusion of the shorter chains results in constraint release for the longer chains, leading to faster motion of the longer chains in the dispersed melt than in a uniform melt. This large variation in mobility of chains within entangled melts offers a means to tune the viscoelasticity of these melts by manipulating chain mobility through dispersity.
One would expect that the frequency dependence rheological response of the viscosity will be strongly affected by the fact that the shortest and longest changes move at significantly faster and slower rates compare with a uniform system.  Studies of this effect are currently on the way.

KMS was supported in part by the National Research Council Associateship Program at the US Naval Research Laboratory. DP kindly acknowledged NSF DMR 1611136 for partial support. This work was supported by the Sandia Laboratory Directed Research and Development Program. This work was performed, in part, at the Center for Integrated Nanotechnologies, an Office of Science User Facility operated for the U.S. Department of
Energy (DOE) Office of Science.  Sandia National Laboratories is a multi-mission laboratory managed and operated by National Technology and Engineering Solutions of Sandia, LLC, a wholly owned subsidiary of Honeywell International, Inc., for the U.S. Department of Energy's National Nuclear Security Administration under Contract DE-NA-0003525.

\end{document}